# Extreme Sub-Wavelength Light Confinement in Plasmonic Film-Coupled Nanostar Resonators

Negar Charchi[1], Ying Li[2], Margaret Huber[1], Elyahb Allie Kwizera[3], Xiaohua Huang[3], Christos Argyropoulos[2], Thang Hoang[1]

[1]Department of Physics and Materials Science, The University of Memphis, Memphis, TN 38152

[2]Department of Electrical and Computer Engineering, University of Nebraska-Lincoln, Lincoln, NE 68588

[3]Department of Chemistry, The University of Memphis, Memphis, TN 38152

Confining light in extreme subwavelength scales is a tantalizing task. In this work, we report a study of individual plasmonic film-coupled nanostar resonators where plasmonic optical modes are trapped in ultrasmall volumes. Individual gold nanostars, separated from a flat gold film by a thin dielectric spacer layer, exhibit a strong light confinement between the sub-10 nm volume of the nanostar's tips and the film. Through dark field scattering measurements of many individual nanostars, a statistical observation of the scattered spectra is obtained and compared with extensive simulation data to reveal the origins of the resonant peaks. We observe that an individual nanostar on a flat gold film can result in a resonant spectrum with single, double or multiple peaks. Further, these resonant peaks are strongly polarized under white light illumination. Our simulation data revealed that the resonant spectrum of an individual film-coupled nanostar resonator is related to the symmetry of the nanostar, as well as the orientation of the nanostar relative to its placement on the gold substrate. Our results demonstrate a simple method to create an ultrasmall mode volume plasmonic platform which could be useful for applications in sensing or enhanced light-matter interactions.

Attempts to create ultrasmall hot spots to confine light at the nanoscale have led to various nanophotonic system designs realized via both lithographic and colloidal approaches [1-9]. Among these, plasmonic nanoparticles have proven to be an excellent platform to strongly confine light at length scales much smaller than the diffraction limit [4, 10-13]. As a general rule of thumb, the mode volume (or the hot spot) of a plasmonic nanoparticle is scaled with the size of the particle. Further, the resonant wavelength is also proportional to the size of the particle itself [4, 10, 11, 14]. For instance, the resonant wavelength of spherical gold (Au) occurs between 550 - 800 nm for particle sizes ranging from 30-100 nm. Similarly, for silver nanocubes with sizes varying from 50-150 nm, their gap mode resonant wavelengths change from around 600 nm to 1200 nm [4, 10, 14]. However, smaller particle sizes (< 30 nm) will limit the capability to perform single particle measurements due to the limited possibility of single particle identification, for example, under a dark field illumination. Therefore, the plasmonic hot spot volumes are usually in the order of a few tens of nanometers due to the relatively large particle sizes. Other lithographic techniques also attempted to create very small mode volume structures [2, 7] but the results are still limited to around 30 nm due to several limitations of conventional electron beam lithography systems. Recently, many works have devoted considerable attention to the optical properties of plasmonic nanostars [15-19], which have several advantageous characteristics. Unlike many other previously studied metallic nanoparticles with given symmetries, such as nanospheres [20, 21] or nanocubes [10, 14], nanopopcorns [22, 23] and nanostars [16-18] exhibit to a certain degree non-uniform distribution of localized plasmonic fields along their spikes or tips. Nanostars have recently attracted intensive attention in nanophotonics [18, 19], biology [24-27], and materials science [5, 28, 29], thanks to their ability to localize electromagnetic fields in tiny hot spots at their highly engineered nanoscale tips. However, it is also noted that the majority of recent studies related to optical properties of Au nanostars have studied ensembles of particles [5, 24, 30], which cannot reveal the true origin of the mode resonances. Several works have also looked into the plasmon resonances of individual star-shaped nanoparticles in free space or in a solution [9, 15, 31]. Shao et. al. have also reported optical properties of single nanostars on an indium tin oxide substrate [17].

Here, we demonstrate an alternative approach to confine plasmonic optical modes in ultrasmall volumes by using film-coupled nanostar resonators. Specifically, a Au nanostar is placed on top of a Au film, separated by a thin (5 nm) dielectric polymer layer. We show that the plasmon resonances of the nanostar's tips are significantly enhanced due to the resulted extreme sub-wavelength mode volume. Our results clearly indicate that the highly intense electromagnetic field enhancement in the tiny gaps formed between the nanostar's tips and the Au film can strongly confine light at a sub-10 nm volume. Moreover, the nanostars are clearly visible under a typical dark field microscope, despite the ultrasmall mode volume, allowing easy identification and measurement of single nanoparticles. Indeed, when illuminated by a white light source, the nanostars that are coupled to a Au film appear to be much brighter than the same particles that are deposited on a glass substrate. The strong light-matter interactions achieved by the proposed plasmonic film-coupled nanostar platform can be used for sensing [13], or nonlinear [32, 33], and quantum [5, 34] optical applications.

The sample structure of this present study consists of colloidal synthesized iron oxide-gold core-shell nanostars with an approximate size of 75 nm placed over a Au film (50 nm thick, fabricated by electron beam evaporation method). The magnetic iron oxide core was about 35 nm and the Au shell layer was 40 nm. The iron oxide-gold core-shell nanostar's original purpose was to make use of its dual-functionality, namely magnetic and plasmonic properties, for biomedical applications [30]. However, for this particular work the iron core's role was to serve as a catalyst for the nanostar growth and its influence on the plasmon resonances of the Au nanostar is negligible. The core-shell Au nanostar and the Au film were separated by a 5 nm polymer spacer layer sandwiched in between (Fig. 1(a)).

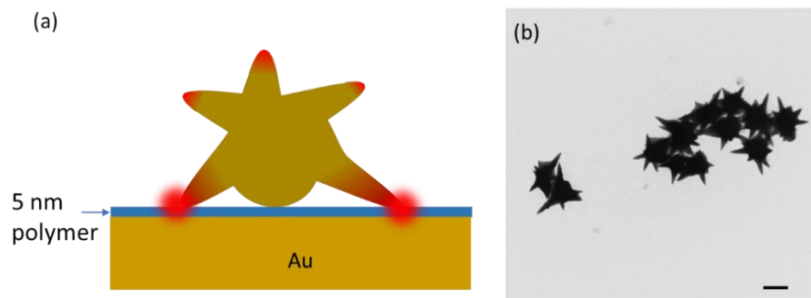

**Fig. 1**: (a) Schematic of the proposed ultrasmall mode volume plasmonic film-coupled nanostar resonator: A nanostar with sub-10 nm tips situated on top of a Au film, separated by a very thin 5 nm polymer dielectric layer. (b) TEM image of typical nanostars. Scale bar is 50 nm.

The polymer spacer layer was formed by alternative dip coating with polyelectrolytes (PE) [3]. Specifically, five alternating positive PAH (poly(allylamine) hydrochloride) and negative PSS (polystyrenesulfonate) layers (1 nm each) were used. The thickness of the spacer layer may affect the resonant wavelength and the amplitude of the scattered light intensity of the coupled nanostar-film resonator, depending on the incident angle of the illuminated light [35]. The spikes (or tips) of the Au nanostars were estimated, via Transmission Electron Microscopy (TEM) image (Fig. 1(b)), to be less than 10 nm. Indeed, the small tips of the nanostar permit the formation of extremely tiny nanoscale hot spots. The deposition technique of nanostars on a Au film was described elsewhere [3] and the surface coverage of nanostars was less than 5% (Fig. 2).

In the experimental procedure, individual nanostars were identified and isolated by a set of lenses and pinholes to allow single particle measurements. Specifically, individual nanostars were illuminated by a white light source through a 100X Nikon objective lens operating in the dark field mode. The scattered light from particles was collected by the same objective, filtered by a pinhole aperture, refocused onto the entrance slit of a spectrometer (Horiba Jobin-Yvon iHR550) and analyzed by a CCD (Charged-Coupled Device) camera (Horiba Jobin-Yvon Synapse). For the TE (transverse electric) and TM (transverse magnetic) polarized light, our customized optical setup allowed for a rotatable linear polarizer to be placed in a microscope slider or on the optical path in free space. The final scattering spectrum of a particular particle was normalized as $Scattering = \frac{I_{NS}-I_{SS}}{I_{WL}-I_{dark}}$ where $I_{NS}, I_{SS}, I_{WL}$ and $I_{dark}$ are the light intensities measured from the nanostar, substrate, white light standard reflection piece, and the CCD dark counts, respectively. For better visualization (Figs. 2(a) and 2 (b)) and for scattering intensity analysis (Fig. 2(c)), the nanostar scattering images were captured by a color camera (Thorlabs DCC 1645 C) and by a high dynamic range black and white camera (Photometrics CoolSnap DYNO).

Figures 2(a) and 2(b) show the dark field scattering color images of Au nanostars for two different cases: on a flat Au film and on a glass substrate, respectively. Individual Au nanostars are visible as bright, red dots. For both cases, an identical 5 nm thin polymer layer was coated on the Au film or glass substrate prior to the nanostar deposition. Due to the random distribution of the nanostars on a given substrate (Au or glass), it is not simple to compare the resonant characteristics of individual nanostars with different polymer layer gap thicknesses. A previous computational study by Solis et al. [35] has shown that depending on the incident angle of the illumination the thickness of the gap may have an effect on the spectral characteristics as well as the amplitude of the resonant peaks. In our optical measurements, because the nanostars' orientations were random, the incident angle of the illumination could vary from one to another. Therefore, throughout this work we used a fixed polymer gap thickness of 5 nm to limit the number of involved parameters. It can be clearly seen that the nanostars on a Au film in Fig. 2(a) are much brighter compared to similar stars on a glass substrate (Fig. 2(b)). This indicates that the nanostars have coupled with the underneath Au film and resulted in a strong plasmonic resonance response. Furthermore, in order to confirm the individual bright scattering dots shown in Figs. 2(a) and 2(b) are from individual nanostars, Fig. 2(c) shows a scanning electron microscope (SEM) image of nanostars on a Au film. Even though the SEM image cannot reveal detail of individual nanostar's tip (inset of Fig. 2(c)), it is clear that the nanoparticles were well separated and did not form clusters. Further, we would like to mention that we could not perform the SEM measurement of nanostars on a glass substrate (without a Au mirror film) due to a charging issue. However, we believe that the distribution of the nanostars on a glass sample is the same as on a Au film sample, because in both cases the surface of the samples were coated with an identical 5 nm PE spacer layer.

The images presented in Figs. 2(a) and 2(b) were captured by a color camera that has a 255-pixel depth for each color (red, green, and blue) and is only sensitive in the visible frequency range. In order to get the true image intensity from near UV to near IR (400-1050), we used another black and white CCD camera to take images of the nanostars for both cases. Fig. 2(c) shows the histogram of the integrated scattering intensity of individual nanostars for both samples. Each intensity value of an individual nanostar was integrated over an area of 20x20 pixels of the camera image. It is clear that the intensity from the film-coupled nanostar resonators is much brighter compared with the nanostars on a glass substrate.

Figure 3 shows the scattering spectra for individual nanostars on a Au film (Figs. 3(a) and 3(b)) and on a glass substrate (Figs. 3(c) and 3(d)). Due to the fact that the deposited nanostars on a surface were randomly distributed, the localized surface plasmon resonances will strongly depend on how an individual nanostar is oriented on the surface [15, 17, 31]. As shown in Fig. 3, our measurements indicate that the resonant spectra of individual nanostars can be categorized into two main groups, those with single (Figs. 3(a) and 3(c)) and double (Figs. 3(b) and 3(d)) resonant peaks. For each sample, i.e. on glass or on Au film substrate, we have measured more than one hundred individual particles to draw this conclusion. During several occasions, we also observed resonant spectra with multiple (more than two) plasmon resonances. Previous studies [9, 15, 31] have also observed similar resonant features for the case of nanostars in air. However, in contrast to these previous works, our case has the existence of the Au film in close proximity to the tips of the nanostars, which helps to drastically enhance the mode confinement at the bottom tips.

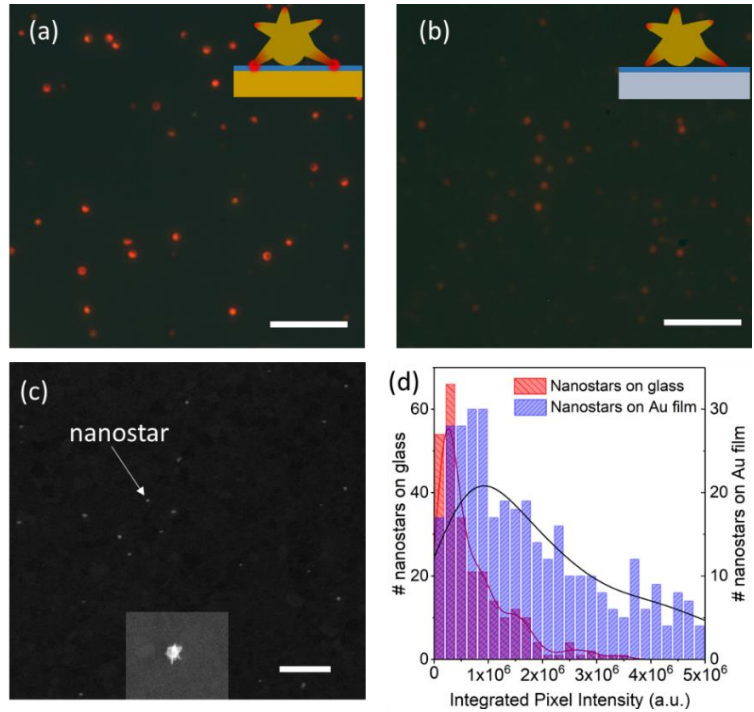

**Fig. 2**: Dark-field scattering and SEM studies. (a) Dark-field image of nanostars on a Au film and (b) on glass substrate. Both (a) and (b) were plotted by using the same brightness scale. (c) SEM image of the nanostars on a Au film substrate. (Scale bars are 1 $\mu m$). (d) Integrated pixels' intensity of the CCD images for nanostars on a Au film (blue) and on a glass substrate (red). Solid curves represent the intensity distributions.

It is also noticed that for the nanostars on a Au film the spectral full width at half maximum (FWHM) of the plasmon resonances was narrower compared with the FWHM for the nanostars on a glass substrate. Specifically, for the nanostars on a Au film with a single resonant peak the average of measured FWHMs was 35.7 nm while for the nanostars on a glass substrate the averaged FWHMs was 55 nm. For the nanostars on a Au film that exhibited double resonant peaks the average of measured FWHMs were 32.5 nm and 32.0 nm for the lower and longer wavelength resonant peaks, respectively. On a glass substrate, these values were 48 nm and 47 nm for the lower and longer wavelength resonant peaks, respectively. As we have mentioned above the spectral characteristics of an individual nanostar resonator may vary depending on the orientation of the nanoparticle and the illumination condition. Nevertheless, our statistical data indicates a strong light confinement by the nanostars on the Au film. Furthermore, the spectral characteristics of the film-coupled Au nanostars presented in our current work are different compared to other film-coupled nanoparticle resonators, such as nanospheres [12] and nanocubes [4, 10], where only a single peak was observed. As we will be discussing below, the appearance of a single peak or double peaks in the resonant spectrum is closely related to the geometry of the nanostar and how the nanostar contacts with the substrate.

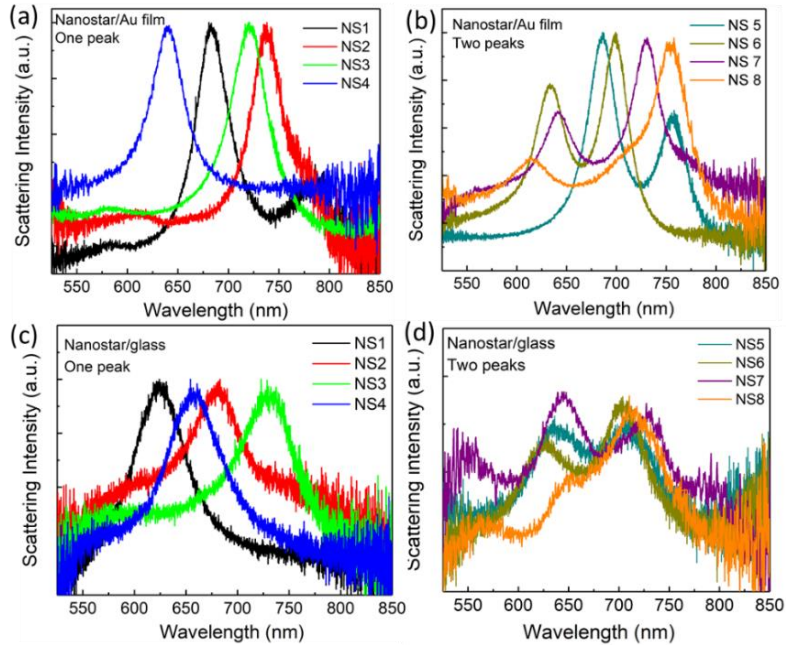

**Fig. 3**: (a)-(b) Measured scattering spectra from individual nanostars on a Au film. Each panel represents a different group of spectral characteristic. (c)-(d) Similar measurements but for nanostars on a glass substrate.

The results presented in Figs. 3(a) and 3(c) may indicate that the measured spectra were from nanostars with poorly formed tips [31]. However, in such a situation one would expect an unpolarized resonant spectrum for a given particle due to its less asymmetrical shape. This is not always the case in our observations. In Fig. 4 we show the polarization dependent measurements of individual nanostars on Au film for two different cases: single and double peaks. In the case of the single peak, the TE and TM polarized components resulted in a degree of polarization of $P = 28.7\%$. Here, TE and TM polarized waves were chosen to be perpendicular to each other and the polarization degree was defined as $P = \frac{|I_{TE}-I_{TM}|}{I_{TE}+I_{TM}}$, where $I_{TE}$ and $I_{TM}$ are the scattering intensities for TE and TM signal, respectively. For the two-peak resonance case, the degrees of polarization were $P = 61\%$ and $P = 86\%$ for the 687 nm and 758 nm resonant peak, respectively. Previous studies by Hrelescu [9], Hao [15], and Nehl [31] have also observed a similar polarization behavior in the individual nanostar's scattering spectrum, which was related to the different tip geometries in these nanoparticles. Thus, these measurements allow us to conclude that the collected signal was from star-shaped nanoparticles.

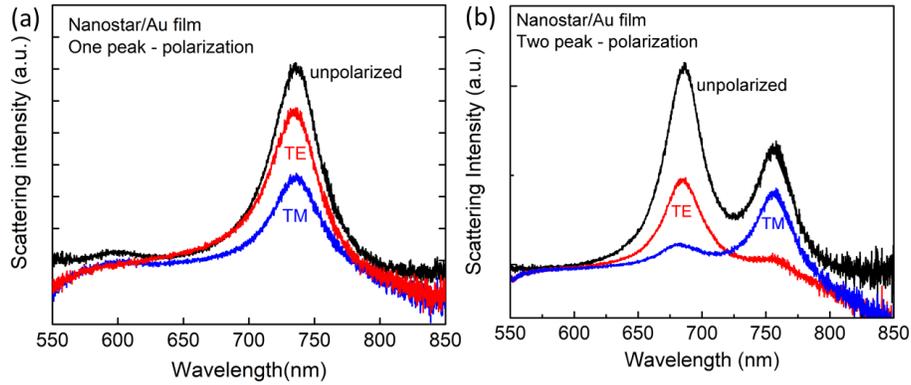

**Fig. 4**: Scattering spectra detected for TE and TM polarization components for (a) single peak and (b) two peaks cases.

The deposited nanostars on a Au substrate can take random positions. In an ideal situation, one can label the position of a nanostar, measure its scattering spectrum and then proceed with a SEM measurement of the same particle [36]. Via this way, one can correlate the geometry of the film coupled nanostar resonator with the optical property of the same particle. However, our current measurement capability does not allow us to follow this approach. Therefore, we must turn our attention to computer simulations. We use the commercial finite-element simulation software Comsol Multiphysics to model several geometries that would represent possible experimental scenarios, including asymmetric properties, number of tips that touch the bottom film, and the polarizations of the scattered light from individual nanostars. Figure 5 shows the simulated extinction spectra for nanostars at several different configurations and geometries placed on a Au film. In order to calculate the extinction cross section (ECS) of the film-coupled nanostars, we employ the scattered-field formulation that directly computes the scattered fields by subtracting the analytical solution of an incident plane wave in the absence of the nanostars, which is considered to be the background field [33]. Perfectly matched layer (PML) boundary conditions are used outside the simulated spherical area to completely truncate the computation domain and absorb the scattered energy. The dimensions and geometries of the nanostar were predicted from the actual shapes of the particles taken from the TEM images, as shown in Fig. 1(b). For simplicity, we first consider a nanostar that has 12 relatively short and equal-length sharp tips branching out from a central spherical core. However, as observed from the experimental results of the TEM measurements, the nanostars' tips were not necessarily uniformly distributed around the bulk of the particle. In our first initial calculations, we assumed that the 12 tips are uniformly distributed around the central core resulting in a star-like dodecahedral gold nanocrystal and such nanostar has only one tip touching the bottom surface. The core diameter is around 75 nm and the tip length is about 30 nm. Each tip has corn shape with small apex (~ 3 nm) and 34°- 52° corn angle. Figure 5(a) shows the polarized (TE and TM) and unpolarized extinction spectra for this specific situation. The dielectric function of Au used in the simulations was taken from experimental data [37]. A thin 5 nm polymer layer was placed in the nanogap with refractive index $n$=1.4, which is identical to the thin polymer layer used in the experiment. The associated field enhancement distributions are shown in Fig. S1 of the Appendix. We find that the symmetric nanostar under TE polarization illumination will not exhibit a plasmonic gap mode inside the thin polymer dielectric layer.

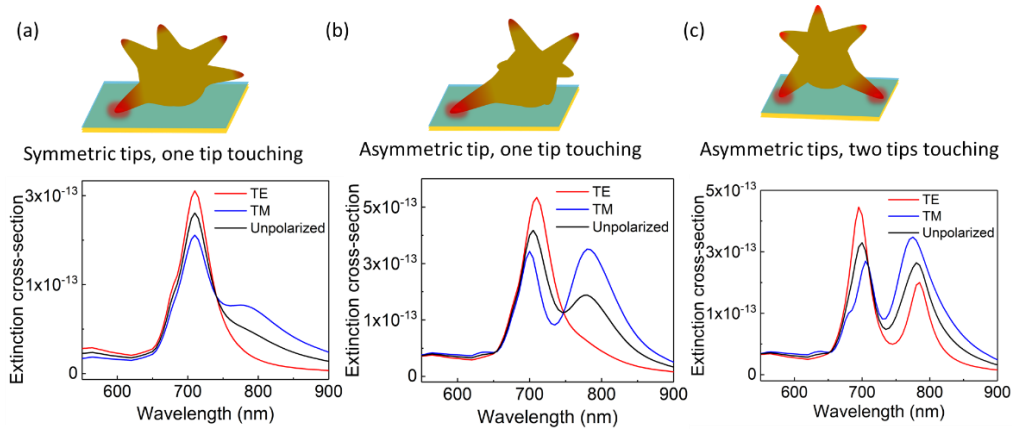

**Fig. 5**: Simulated extinction spectra for three different geometries as schematically shown in the top row: (a) A single equal-length tip touches the substrate, (b) one asymmetric tip touches the substrate, and (c) two asymmetric tips touch the bottom substrate.

The result of this particular simulation matches well with the measurement scattering spectra presented in Figs. 3(a) and 4(a). The TE and TM polarized waves in the simulation were perpendicular to each other. After checking the field enhancement distribution for each polarization illumination, we find that a strong plasmonic gap mode can be excited only under TM polarization illumination and no gap enhancement exists with TE polarization incidence. In addition, the incidence angle for this case is 45°, because the symmetric nanostar under normal illumination will not exhibit a plasmonic gap mode inside the thin (5 nm) polymer dielectric layer. This has also been demonstrated in reference [35]. However, after we introduce some asymmetry in the nanostructure, such as elongating either one or two of the asymmetric tips, a strong gap field enhancement can be achieved for both TE and TM polarization even under normal incidence illumination. We also study the cases when the nanostar has one and two asymmetric tips touching the surface under a normal incident illumination. For these scenarios, we observed two distinct peaks in the resonant spectra, which are indeed polarized differently for TE and TM illuminations. These later configurations result in an excellent agreement with the measured data presented in Figs. 3(b) and 4(b). In addition, we also performed the simulation for similar configurations of nanostars on Au film at different thicknesses of the polymer layer (not shown here). We also note that there is a small wavelength offset between the computational data and the experimental data shown in Figs. 3 and 4. Indeed, in the experimental data there was quite a broad range of resonant wavelengths observed due to the fact that the nanostars were randomly distributed as well as difference in size. Additional simulation data regarding this wavelength offset is shown in Fig. S2 of the Appendix. Therefore, the simulation data shown in Fig. 5 demonstrates the polarization properties of these film-coupled nanostars, which are indeed in agreement with the experimental data.

TABLE 1. Estimated mode volume $V_{eff}$ of the plasmonic resonance modes for a single nanostar placed on a Au film corresponding to three different geometries shown in Figs. 5(a)-(c).

| $V_{eff}$ (nm³) | (a) | (b) | (c) |
|---|---|---|---|
| **TE** | $2.30 \times 10^{-5} \lambda^3$ | $1.62 \times 10^{-5} \lambda^3$ | $1.05 \times 10^{-7} \lambda^3$ |
| **TM** | $2.46 \times 10^{-6} \lambda^3$ | $1.81 \times 10^{-5} \lambda^3$ | $9.75 \times 10^{-7} \lambda^3$ |

We also compute and present in Table 1 the estimated mode volume $V_{eff}$ of the plasmonic resonance modes in the three different geometries shown in Figs. 5(a) - 5(c). The mode volume is defined as $V_{eff} = \int \epsilon(\mathbf{r}) |\mathbf{E}(\mathbf{r})|^2 d^3\mathbf{r} / \text{Max}[\epsilon(\mathbf{r})|\mathbf{E}(\mathbf{r})|^2]$, where $\epsilon(\mathbf{r})$ is the material permittivity and the integration is performed over and around the mode's bright spot [38]. From the above equation, the mode volume is characterized as the ratio of total electric energy to the maximum value of the electric energy density, where the maximum electric energy densities in our system are located around the tiny gaps formed between the nanostar's tips and Au film. The results in Table 1 show that a sub-10 nm mode volume can be achieved for both TE and TM polarization illuminations, especially in the case of the nanostar geometry with two asymmetric tips touching the bottom substrate (Table 1(c) and Fig. 5(c)). Compared to the configurations of one tip touching, the maximum electric filed enhancement is increased in the case of two tips touching. Several previous works have also demonstrated very small mode volumes for both plasmonic and dielectric platforms. For instance, Huang et al.[1] have reported a computational work to demonstrate a nanoparticle – on – mirror structure that offers a mode volume below $10^{-7}\lambda^3$. Several other experimental works have also demonstrated $10^{-5}\lambda^3$ plasmonic mode volume [2, 6]. Our method provides a quick approach to create ultrasmall mode volume (considering the case of Fig. 5(c)) plasmonic nanocavities without need for complicated electron lithography procedure.

In conclusion, we have investigated a film-coupled nanostar resonator system that offers extreme sub-wavelength mode volumes. It is demonstrated that by placing a Au nanostar on a flat Au film, ultrasmall mode volumes are formed and strong light confinement is achieved in between the nanostar's tip and the film region. We have performed simulations to correlate the plasmonic resonant features with the geometry of the resonators by considering the asymmetric shape of the nanostar and how it may orient with respect to the underneath film. The film-coupled nanostar resonators provide a new approach to achieve strong light confinement within sub-10 nm mode volumes, which makes them an ideal platform for applications requiring enhanced light-matter interactions at low dimensions.

**Appendix: Simulation of film-coupled nanostar resonators**

For simplicity, we first simulate a symmetric nanostar, containing 12 equal-length sharp tips branching out from a central spherical core (Fig. S1(a)). The 12 tips are uniformly distributed around the central core resulting in a star-like dodecahedral gold nanocrystal. The core diameter is around 75 nm and the tip length is about 30 nm. Each tip has corn shape with small apex (3 nm). We simulate the interaction of such nanostar with a planer Au surface when the nanostar is illuminated at 45° incidence angle under both TE and TM polarization. The separation between nanostar and Au film is due to a thin 5 nm polymer layer with refractive index n = 1.4, which is identical to the thin polymer layer used in the experiment. The associated field enhancement distributions are shown in Figs. S1(a1) and S1(a2). We find that the symmetric nanostar under TE polarization illumination will not exhibit a plasmonic gap mode inside the thin polymer dielectric layer. In order to produce stronger gap plasmonic mode enhancement, we further simulate several additional asymmetric nanostar designs, such as elongating one or two of the tips of the nanostars that will resemble the experimentally used nanostars (Fig. 1(b) in the main text). As shown in Fig. 5(c) of the main text, we simulate the extinction spectra when a nanostar has one asymmetric tip and two tips touching the polymer layer surface. The associated electric filed distribution is shown in Fig. S2(b) under different (TE/TM) polarization normal illumination.

Figures S1(a1) and S1(a2) show the normalized electric field distribution in the case of symmetric tips in the one tip touching geometry under different polarization illumination (TE or TM) by using 45° incidence angle. It is clear that a strong plasmonic gap mode can be excited only under TM polarization illumination. However, after we introduce some asymmetry in the structure, such as by using one or two asymmetric tips, a strong gap field enhancement can be achieved in both TE and TM polarization illumination. Figs. S1(b1)- S1(b4) show four different field enhancement distributions in the case of one asymmetric tip, two tips touching nanostar, with geometry shown in Fig. 5(c) in the main text, where Figs. S1(b1) and S1(b2) correspond to two ECS peaks when the nanostar is illuminated by TE polarization, Figs. S1(b3) and S1(b4) correspond to other two resonance peaks under TM polarization. We find that the value of the field enhancement nearly doubles for the case of asymmetric tips geometry even under normal incidence illumination (Fig. S1(b)), compared to the field enhancement using only symmetric nanostar (Fig. S 1(a)). Furthermore, we note that, around the second resonant wavelength (~780 nm), a strong plasmonic gap mode always exists under both TE (Fig. S1(b2)) and TM (Fig. S1 (b4)) polarization incidence, which is different from the symmetric nanostar response.

A complete statistical analysis by using simulations is not feasible because multiple 3D simulations are required and that will take a very long time to run. However, we present some results of a parametric study by changing the dimensions or shape (one tip or two tips touching or symmetric/asymmetric designs) of the nanostar as shown in Fig. S2. The simulated extinction spectra for different sizes of nanostar match well with the measured scattering spectra presented in Fig. 3 in the main text, where some nanostars have single resonant peak (Fig. S2(a)) and others have double resonant peaks (Fig. S2(b)).

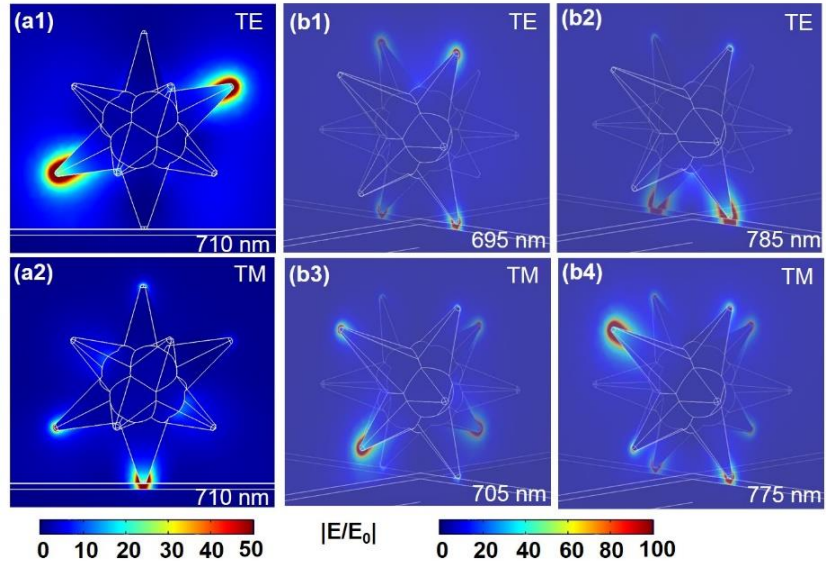

**Fig. S1**. Normalized field distribution for two different geometries with different resonant wavelengths. (a1) - (a2) correspond to the geometry of symmetric tips, i.e., one tip touching nanostar, and (b1) - (b4) correspond to the case of asymmetric tip, i.e., two tips touching nanostar.

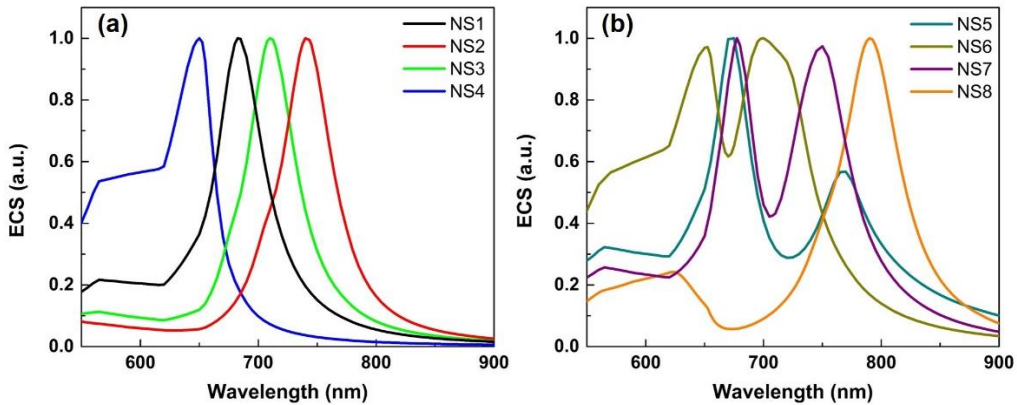

**Fig. S2**. Simulated extinction spectra for several different geometries and shapes of nanostars located on a Au film. (a) Single resonant peak and (b) two resonant peaks cases.

This work was supported by the National Science Foundation (NSF) (Grant # DMR-1709612). TH acknowledges the Ralph E. Powe Junior Faculty Enhancement Award from Oak Ridge Associated

Universities and FedEx Institute of Technology at the University of Memphis. XH acknowledge the support from the National Institutes of Health (Grant No. 1R15 CA 195509-01) and the Fedex Institute of Technology at the University of Memphis.

**References**

1. S. Huang, T. Ming, Y. Lin, X. Ling, Q. Ruan, T. Palacios, J. Wang, M. Dresselhaus, and J. Kong, "Ultrasmall Mode Volumes in Plasmonic Cavities of Nanoparticle-On-Mirror Structures," Small **12**(37), 5190-5199 (2016).

2. M. Kuttge, F. J. García de Abajo, and A. Polman, "Ultrasmall Mode Volume Plasmonic Nanodisk Resonators," Nano Lett. **10**(5), 1537-1541 (2010).

3. T. B. Hoang, Huang, J., Mikkelsen, M. H., "Colloidal Synthesis of Nanopatch Antennas for Applications in Plasmonics and Nanophotonics," J. Vis. Exp. **111**(e53876 (2016).

4. T. B. Hoang, G. M. Akselrod, C. Argyropoulos, J. Huang, D. R. Smith, and M. H. Mikkelsen, "Ultrafast spontaneous emission source using plasmonic nanoantennas," Nat. Commun. **6**(7788 (2015).

5. H. Asgar, L. Jacob, and T. B. Hoang, "Fast spontaneous emission and high Förster resonance energy transfer rate in hybrid organic/inorganic plasmonic nanostructures," J. Appl. Phys **124**(10), 103105 (2018).

6. H. Choi, M. Heuck, and D. Englund, "Self-Similar Nanocavity Design with Ultrasmall Mode Volume for Single-Photon Nonlinearities," Phys. Rev. Lett. **118**(22), 223605 (2017).

7. A. Kinkhabwala, Z. Yu, S. Fan, Y. Avlasevich, K. Müllen, and W. E. Moerner, "Large single-molecule fluorescence enhancements produced by a bowtie nanoantenna," Nat. Photonics **3**(654 (2009).

8. E. B. Ureña, M. P. Kreuzer, S. Itzhakov, H. Rigneault, R. Quidant, D. Oron, and J. Wenger, "Excitation Enhancement of a Quantum Dot Coupled to a Plasmonic Antenna," Adv. Mater. **24**(44), OP314-OP320 (2012).

9. C. Hrelescu, T. K. Sau, A. L. Rogach, F. Jäckel, and J. Feldmann, "Single gold nanostars enhance Raman scattering," Appl. Phys. Lett. **94**(15), 153113 (2009).

10. J. B. Lassiter, F. McGuire, J. J. Mock, C. Ciracì, R. T. Hill, B. J. Wiley, A. Chilkoti, and D. R. Smith, "Plasmonic Waveguide Modes of Film-Coupled Metallic Nanocubes," Nano Lett. **13**(12), 5866-5872 (2013).

11. C. Ciracì, R. T. Hill, J. J. Mock, Y. Urzhumov, A. I. Fernández-Domínguez, S. A. Maier, J. B. Pendry, A. Chilkoti, and D. R. Smith, "Probing the Ultimate Limits of Plasmonic Enhancement," Science **337**(6098), 1072 (2012).

12. J. J. Mock, R. T. Hill, A. Degiron, S. Zauscher, A. Chilkoti, and D. R. Smith, "Distance-Dependent Plasmon Resonant Coupling between a Gold Nanoparticle and Gold Film," Nano Lett. **8**(8), 2245-2252 (2008).

13. J. J. Baumberg, J. Aizpurua, M. H. Mikkelsen, and D. R. Smith, "Extreme nanophotonics from ultrathin metallic gaps," Nat. Mater. **18**(7), 668-678 (2019).

14. A. Moreau, C. Ciracì, J. J. Mock, R. T. Hill, Q. Wang, B. J. Wiley, A. Chilkoti, and D. R. Smith, "Controlled-reflectance surfaces with film-coupled colloidal nanoantennas," Nature **492**(86 (2012).

15. F. Hao, C. L. Nehl, J. H. Hafner, and P. Nordlander, "Plasmon Resonances of a Gold Nanostar," Nano Lett. **7**(3), 729-732 (2007).

16. K. Pandian Senthil, P.-S. Isabel, R.-G. Benito, F. J. G. d. Abajo, and M. L.-M. Luis, "High-yield synthesis and optical response of gold nanostars," Nanotechnolog **19**(1), 015606 (2008).

17. L. Shao, A. S. Susha, L. S. Cheung, T. K. Sau, A. L. Rogach, and J. Wang, "Plasmonic Properties of Single Multispiked Gold Nanostars: Correlating Modeling with Experiments," Langmuir **28**(24), 8979-8984 (2012).

18. J. Ziegler, C. Wörister, C. Vidal, C. Hrelescu, and T. A. Klar, "Plasmonic Nanostars as Efficient Broadband Scatterers for Random Lasing," ACS Photonics **3**(6), 919-923 (2016).

19. A. Kedia and P. S. Kumar, "Gold nanostars reshaping and plasmon tuning mechanism," AIP Conf. Proc. **1512**(1), 232-233 (2013).

20. F. Tam, G. P. Goodrich, B. R. Johnson, and N. J. Halas, "Plasmonic Enhancement of Molecular Fluorescence," Nano Lett. **7**(2), 496-501 (2007).

21. M. Hu, J. Chen, Z.-Y. Li, L. Au, G. V. Hartland, X. Li, M. Marquez, and Y. Xia, "Gold nanostructures: engineering their plasmonic properties for biomedical applications," Chem. Soc. Rev. **35**(11), 1084-1094 (2006).

22. Z. Fan, D. Senapati, A. Khan Sadia, K. Singh Anant, A. Hamme, B. Yust, D. Sardar, and C. Ray Paresh, "Popcorn-Shaped Magnetic Core–Plasmonic Shell Multifunctional Nanoparticles for the Targeted Magnetic Separation and Enrichment, Label-Free SERS Imaging, and Photothermal Destruction of Multidrug-Resistant Bacteria," Chem.: Eur. J. **19**(8), 2839-2847 (2013).

23. W. Lu, A. K. Singh, S. A. Khan, D. Senapati, H. Yu, and P. C. Ray, "Gold Nano-Popcorn-Based Targeted Diagnosis, Nanotherapy Treatment, and In Situ Monitoring of Photothermal Therapy Response of Prostate Cancer Cells Using Surface-Enhanced Raman Spectroscopy," J. Am. Chem. Soc. **132**(51), 18103-18114 (2010).

24. Y. Liu, H. Yuan, R. F. Kersey, K. J. Register, C. M. Parrott, and T. Vo-Dinh, "Plasmonic Gold Nanostars for Multi-Modality Sensing and Diagnostics," Sensors **15**(2)(2015).

25. V. Raghavan, C. O'Flatharta, R. Dwyer, A. Breathnach, H. Zafar, P. Dockery, A. Wheatley, I. Keogh, M. Leahy, and M. Olivo, "Dual plasmonic gold nanostars for photoacoustic imaging and photothermal therapy," Nanomedicine **12**(5), 457-471 (2017).

26. I. G. Theodorou, Q. Jiang, L. Malms, X. Xie, R. C. Coombes, E. O. Aboagye, A. E. Porter, M. P. Ryan, and F. Xie, "Fluorescence enhancement from single gold nanostars: towards ultra-bright emission in the first and second near-infrared biological windows," Nanoscale **10**(33), 15854-15864 (2018).

27. S. K. Dondapati, T. K. Sau, C. Hrelescu, T. A. Klar, F. D. Stefani, and J. Feldmann, "Label-free Biosensing Based on Single Gold Nanostars as Plasmonic Transducers," ACS Nano **4**(11), 6318-6322 (2010).

28. X.-L. Liu, J.-H. Wang, S. Liang, D.-J. Yang, F. Nan, S.-J. Ding, L. Zhou, Z.-H. Hao, and Q.-Q. Wang, "Tuning Plasmon Resonance of Gold Nanostars for Enhancements of Nonlinear Optical Response and Raman Scattering," J. Phys. Chem. C **118**(18), 9659-9664 (2014).

29. B. Munkhbat, J. Ziegler, H. Pöhl, C. Wörister, D. Sivun, M. C. Scharber, T. A. Klar, and C. Hrelescu, "Hybrid Multilayered Plasmonic Nanostars for Coherent Random Lasing," J. Phys. Chem. C **120**(41), 23707-23715 (2016).

30. E. A. Kwizera, E. Chaffin, X. Shen, J. Chen, Q. Zou, Z. Wu, Z. Gai, S. Bhana, R. O'Connor, L. Wang, H. Adhikari, S. R. Mishra, Y. Wang, and X. Huang, "Size- and Shape-Controlled Synthesis and Properties of Magnetic–Plasmonic Core–Shell Nanoparticles," J. Phys. Chem. C **120**(19), 10530-10546 (2016).

31. C. L. Nehl, H. Liao, and J. H. Hafner, "Optical Properties of Star-Shaped Gold Nanoparticles," Nano Lett. **6**(4), 683-688 (2006).

32. Z. Huang, A. Baron, S. Larouche, C. Argyropoulos, and D. R. Smith, "Optical bistability with film-coupled metasurfaces," Optics Letters **40**(23), 5638-5641 (2015).

33. C. Argyropoulos, C. Ciracì, and D. R. Smith, "Enhanced optical bistability with film-coupled plasmonic nanocubes," Appl. Phys. Lett. **104**(6), 063108 (2014).

34. T. B. Hoang, G. M. Akselrod, and M. H. Mikkelsen, "Ultrafast Room-Temperature Single Photon Emission from Quantum Dots Coupled to Plasmonic Nanocavities," Nano Lett. **16**(1), 270-275 (2016).

35. D. M. Solís, J. M. Taboada, F. Obelleiro, L. M. Liz-Marzán, and F. J. García de Abajo, "Toward Ultimate Nanoplasmonics Modeling," ACS Nano **8**(8), 7559-7570 (2014).

36. A. Rose, T. B. Hoang, F. McGuire, J. J. Mock, C. Ciracì, D. R. Smith, and M. H. Mikkelsen, "Control of Radiative Processes Using Tunable Plasmonic Nanopatch Antennas," Nano Lett. **14**(8), 4797-4802 (2014).

37. E. D. Palik, *Handbook of optical constants of solids* (Academic, 1985).

38. B. Bahari, R. Tellez-Limon, and B. Kante, "Directive and enhanced spontaneous emission using shifted cubes nanoantenna," J. Appl. Phys **120**(9), 093106 (2016).